\documentclass{article}
\usepackage{bm,amsmath,amssymb,setspace}
\usepackage{graphicx}
\usepackage{color}

\begin{document}
\doublespacing

\section{Title:} 
Variable slip coefficient in binary lattice Boltzmann models\\
\section{Author:}
Lajos Szalm\'as\\
\section{Address:} 
Department of Physics, University of Rome "La Sapienza", Piazzale Aldo Moro 2, Rome 00185, Italy\\
\section{Email:}lszalmas@gmail.com\\

\section{Abstract} 
We present a new method in order to get variable slip coefficient in 
binary lattice Boltzmann models to simulate gaseous flows. Boundary layer theory is 
presented. We study both the single- and multi-fluid BGK-type models as well. The 
boundary slip and the Knudsen layer are analyzed in detail. Benchmark simulations 
are carried out in order to compare the analytical derivation with the numerical 
results. Excellent agreement is found between the two situations.\\
\\
\section{Keywords:} 
lattice Boltzmann, binary mixtures, slip coefficient, microflows, Knudsen number\\
\\
\section{PACS:} 
05.20.Dd,47.11.-j,47.45.Ab\\


\section{Introduction}

Over the last decades, the lattice Boltzmann model (LB) has become a versatile tool in modeling 
complex hydrodynamic problems \cite{ref1}. Recently, it has attracted considerable attention in 
modeling rarefied gas flows. These flows are important in several applications ranging 
from micro-electromechanical systems (MEMS) \cite{ref1a} to aeronautical applications.
With increasing rarefaction characterized by the Knudsen number, the ratio of the molecular mean-
free path and the macroscopic sizes, the flow experiences some interesting phenomena, such as the 
gaseous slip along the solid wall. The slip phenomenon is one of the cornerstone of micro-channel 
applications. In LB, several methods have been proposed for the description of the slip flow,
mostly based on kinetic boundary conditions \cite{ref2,ref3,ref4,ref5,ref6,ref7,ref8,ref8a}, and 
analytical formulas have been derived for the slip velocity and the slip coefficient. {
We mention that the first kinetic boundary condition was introduced by Lim et al. \cite{ref8a}.} 
The above approaches focused on one-component gases and the slip phenomenon in mixture LB models is yet 
not understood. In addition, it is worth mentioning that the bare diffuse reflection
boundary condition can not reproduce the correct value of the slip coefficient \cite{ref11}.

In this paper, we present a boundary layer theory for binary lattice Boltzmann models and 
introduce a new boundary treatment in order to get tunable slip coefficient in the method. 
We consider both single and multicomponent BGK-type collision operators as well. In the 
framework of the boundary layer theory, analytical formula is derived for the slip coefficient. 
Simulations are performed to validate the model. The analytical results are in excellent 
agreement with the ones obtained from the numerical simulations.

\section{The binary mixture model}

We begin our consideration with the binary discrete Boltzmann equation for single- and multi-fluid 
BGK-type collision operators
\begin{eqnarray}
\label{1}
\partial_t f^\sigma_i(\bm{x},t) + c_{\sigma ia} \partial_a f^\sigma_i(\bm{x},t)&=&
(\nu_1+\nu_m)[ f^{\sigma(e)}_i(\rho_\sigma,\bm{u}_\sigma)-f^\sigma_i]+\nonumber\\
&&\nu_m[f^{\sigma (e)}(\rho_\sigma,\bm{u})-f_i^{\sigma (e)}(\rho_\sigma,\bm{u}_\sigma)],
\end{eqnarray}
where $f^\sigma_i,f_i^{\sigma (e)}$ denote the single-specie distribution function and equilibrium 
distribution function, $c_{\sigma ia}$ are the discrete speed vectors and $\nu_1,\nu_m$ are
the collision frequencies related to the transport coefficients. 

The macroscopic quantities
are obtained as the moments of $f^\sigma_i$ as
\begin{eqnarray}
\label{2}
n_\sigma=\sum_k f^\sigma_k, \quad n_\sigma \bm{u}_\sigma=\sum_k f^\sigma_k \bm{c}_{\sigma k},
\end{eqnarray}
where $n_\sigma,\bm{u}_\sigma$ are the particle density and velocity, respectively. 
Further, 
\begin{eqnarray}
\label{3}
\rho_\sigma=n_\sigma m_\sigma,
\end{eqnarray}
where $\rho_\sigma,m_\sigma$ are the specie mass density and the specie mass, respectively.
The mixture macroscopic quantities are defined by
\begin{eqnarray}
\label{4}
 \quad n=\sum_\sigma n_\sigma, \quad \rho=\sum_\sigma \rho_\sigma, 
\quad \rho \bm u=\sum_\sigma \rho_\sigma \bm{u}_\sigma,
\end{eqnarray}
where $n,\rho,\bm{u}$ denote the mixture particle density, mixture mass density and 
mass-averaged mixture velocity, respectively. The generic equilibrium distribution
function is chosen as the second order truncation of the Maxwellian
\begin{eqnarray}
\label{5}
f^{\sigma (e)}_i( \rho_\sigma,\bm{u})=w_i \frac{\rho_\sigma}{m_\sigma} \left[ 1+\frac{c_{\sigma ia} u_{a}}{c_\sigma^2}+
\frac{(c_{ \sigma ia}c_{ \sigma ib}-c_\sigma ^2\delta_{ab})u_{a}u_{b}}{2 c_\sigma^4} \right].
\end{eqnarray}

We work in two dimensions, and the discrete speed vectors are given by $\bm{c}_{\sigma i}=c_\sigma \bm{e}_i$,
where $\bm{e}_i$ is the defined as the D2Q9 non-dimensional velocity model
\begin{eqnarray}
\bm e_{i}=\left\{
\begin{aligned}
&(0,0) && \mbox{for} \quad i=0\\
&\left(\sqrt{3}\cos(\frac{\pi}{2}i-\frac{\pi}{2}),\sqrt{3}\sin(\frac{\pi}{2}i-\frac{\pi}{2})\right) && \mbox{for} \quad i=1\dots 4\\
&\left(\sqrt{6}\cos(\frac{\pi}{2}i-\frac{\pi}{4}),\sqrt{6}\sin(\frac{\pi}{2}i-\frac{\pi}{4})\right) && \mbox{for} \quad i=5\dots 8.
\label{4a}
\end{aligned}
\right.
\end{eqnarray}
The corresponding weights are defined by
\begin{eqnarray}
w_{i}=\left\{
\begin{aligned}
&4/9 && \mbox{for} \quad i=0\\
&1/9 && \mbox{for} \quad i=1\dots 4\\
&1/36 && \mbox{for} \quad i=5\dots 8.
\label{4b}
\end{aligned}
\right.
\end{eqnarray}
{
Note that this velocity model is different from the original nine-speed model 
\cite{ref16}. The velocity vectors are rescaled with the $\sqrt{3}$ multiplier. This is 
a convenient choice for our purposes and our calculations. The definition of the model is given 
by the above equations, Eq. (\ref{4a},\ref{4b}). To complete our definition, we mention
that the non-dimensional sound speed of the model is unity.}
The specie sound speed is given by $c_\sigma=c\sqrt{m/m_\sigma}$, where $m=\rho/n$ is the 
averaged mass and $c$ is the mixture sound speed. 

Before going to the boundary layer theory, some comments are in order. The collision operator 
in the governing equation, Eq. (\ref{1}), is equivalent with the so-called Hamel model \cite{ref10}, 
which describes coupled relaxation towards the local specie and mixture equilibriums \cite{ref11,ref12}. With the choice 
of $\nu_1=0$, the collision model results in the single-fluid BGK model. In the two-fluid model, the mixture
viscosity is given by $\mu=\rho c^2/ (\nu_1+\nu_m)$ and the diffusivity takes the value of 
$D=m^2 c^2/(m_1 m_2 \nu_m)$.

\section{Boundary layer theory}

We turn our attention to the half space problem experienced by a gas flow along a solid
wall. This is the so-called Kramers problem in kinetic theory \cite{ref12a}. The solid wall lies in the
$y$ direction and is located at $x=0$. In the $x>0$ region, the gas medium exhibits a shear flow.
Under these circumstances, we are looking for the macroscopic velocity obtained from Eq. (\ref{1}).
This problem can be simplified by introducing a new reduced distribution function instead of
$f_i^\sigma$
\begin{eqnarray}
\label{6}
F_1^\sigma&=&m_\sigma c_\sigma \sqrt{3} [f^\sigma_6-f^\sigma_7],\nonumber\\
F_0^\sigma&=&m_\sigma c_\sigma \sqrt{3} [f^\sigma_2-f^\sigma_4],\nonumber\\
F_2^\sigma&=&m_\sigma c_\sigma \sqrt{3} [f^\sigma_5-f^\sigma_8].
\end{eqnarray}
We also introduce the following weights, $\omega_i$, and one-dimensional velocity vectors, $\epsilon_i$,
corresponding to the new distribution function in such a way that
\begin{eqnarray}
\label{7a}
&\omega_1=1/6,\quad \omega_0=2/3, \quad \omega_2=1/6,\\
\label{7b}
&\epsilon_1=-\sqrt{3}, \quad \epsilon_0=0, \quad \epsilon_2=\sqrt{3}.
\end{eqnarray}
With using the reduced distribution function, the macroscopic specie velocity is obtained by 
$\rho_\sigma u_{\sigma y}=\sum_{k} F_k^\sigma$.

The reduced distribution function obeys the following governing equation 
obtained from Eq. (\ref{1}) in the steady state
\begin{eqnarray}
\label{8a}
c_\sigma \epsilon_i \partial_x F_i^\sigma&=&\nu_1\omega_i\sum_k F_k^\sigma+\nu_m\omega_i\frac{\rho_\sigma}{\rho}\sum_k \sum_{\sigma'} F_k^{\sigma'}-(\nu_1+\nu_m)F_i^\sigma.
\end{eqnarray}
This equation is the boundary layer equation, which needs to be solved for the macroscopic 
velocity. The solution of this differential equation is obtained in the form $F_i^\sigma=F_i^{\sigma (0)} \exp(\lambda x)$.
As a consequence, we obtain the following generalized eigenvalue problem 
\begin{eqnarray}
\label{9a}
c_\sigma \epsilon_i \lambda F_i^{\sigma (0)}&=&\nu_1\omega_i\sum_k F_k^{\sigma (0)}+\nu_m\omega_i\frac{\rho_\sigma}{\rho}\sum_k \sum_{\sigma'} F_k^{\sigma' (0)}-(\nu_1+\nu_m)F_i^{\sigma (0)}.
\end{eqnarray}
This equation is written out in components in the Appendix, Eq. (\ref{9}). The eigenvalues of the problem 
are obtained by $\lambda=[0,0,\lambda_0,-\lambda_0]$, where $\lambda_0$ is listed in the Appendix, Eq. (\ref{10}).
The two zero eigenvalues correspond to a linear flow profile, while the other two to a growing and a 
decaying exponential solution. We remind that in the single component D2Q9 models
the non-zero eigenvalues are absent. However, these two solutions in the mixture case describe the Knudsen 
layer function fading away in order of the mean free path far from the wall.

The linear flow profile corresponds to the hydrodynamic shear flow in the bulk region. After some straightforward 
algebra, one can obtain the linear solution from Eq. (\ref{8a}) as
\begin{eqnarray}
\label{11}
F^{\sigma(H)}_i=\omega_i \rho_\sigma u^{(H)}_y-\omega_i \frac{c_\sigma \epsilon_i}{\nu_1+\nu_m} \rho_\sigma \frac{\partial u^{(H)}_y}{\partial x},
\end{eqnarray}
where the $H$ superscript denotes the hydrodynamic part of the distribution 
function and the velocity.

The general solution of the boundary layer equation, Eq. (\ref{8a}), is given by the linear combination
of the hydrodynamic shear solution, $F^{\sigma(H)}$, and a Knudsen layer part corresponding to the
nonzero eigenvalues. {
Here, we mention that in our work, the Knudsen layer concept is 
identical with the boundary layer provided by the 
established model and obtained from the analytical derivation. This is because the LB model 
is a special discrete ordinate method for the solution of the Boltzmann equation \cite{ref17,ref18}. 
As it has been recently shown, Knudsen layer appears in discrete kinetic models
once the reflected populations have different $c_{ix}$ normal components \cite{ref18}. 
As a result, the boundary layer appearing }{
in the model is physically relevant. However, in 
order to obtain quantitatively correct information, one needs to 
apply optimized boundary condition presented below.}

Before determining the slip coefficient in the model, we theoretically show that the shear stress is 
independent of the Knudsen layer part. Indeed, the shear stress obtained from the
Knudsen layer solution is given by $P^K_{xy}=\sum_k \sum_\sigma m_\sigma c_{\sigma kx}c_{\sigma ky} f_k^\sigma
=\sum_i \sum_\sigma c_\sigma \epsilon_i F^{\sigma(0)}_i \exp (\lambda x)$.
It is quickly realized that $P^K_{xy}$ is always zero for $\lambda \neq 0$, because 
$\lambda \sum_i \sum_\sigma c_\sigma \epsilon_i F^{\sigma (0)}_i$ is nothing just the sum of the left 
hand side of all equations of Eq. (\ref{9}), which is identically zero because the 
right hand side becomes the collision invariant of the momenta for the overall mixture.
This is an important property of the half space problem and remains valid for other velocity 
models as well. As a result, the shear stress at the wall can be 
obtained from the hydrodynamic solution, Eq. (\ref{11}), by
\begin{eqnarray}
\label{13}
P_{xy}=\sum_i \sum_\sigma c_\sigma \epsilon_i F^{\sigma(H)}_i=-\rho \frac{c^2}{\nu_1+\nu_m} \frac{\partial u^{(H)}_y}{\partial x}.
\end{eqnarray}
In this way, the velocity gradient is always available from the shear stress independently
of the unknown Knudsen layer function. 

\section{Variable slip coefficient}

In rarefied gas flows, the slip velocity, $u_s$, is considered as the \textit{extrapolated} gas 
velocity at the wall without the Knudsen layer. It is given in the following form
\begin{eqnarray}
\label{15}
u_s=\alpha_v \frac{\mu v_0}{P} \frac{\partial u^{(H)}_y}{\partial x},
\end{eqnarray}
where $\alpha_v,\mu,v_0,P$ are the slip coefficient, the viscosity, the mixture reference 
speed and the pressure, respectively. In the present model, $v_0=\sqrt{2} c$ and $P=\rho c^2$. 
{
In Eq. (\ref{15}),  $\alpha_v$ is the so-called viscosity based slip coefficient 
\cite{ref15a,ref15b}, which is a useful definition, since it connects the slip coefficient to a 
transport coefficient, which is always available independently of the molecular details of the 
rarefied gas.}
In order to get the slip coefficient, we need to determine the magnitude of the hydrodynamic solution in 
the boundary layer. This is fixed by the used boundary condition at the wall.

We introduce a modified diffuse boundary condition in the model in order to get variable
slip coefficient. The boundary treatment for the reflected populations, $e_{ix}>0$, are 
written by
\begin{eqnarray}
\label{16}
f_i^\sigma=f_i^{\sigma(e)}(\rho_\sigma,\bm{0})+(1-a) w_i \frac{c_{\sigma ix} c_{\sigma iy}}{c^4} \frac{\rho_\sigma}{\rho m} P_{xy},
\end{eqnarray}
where $a$ is a yet unknown parameter, which will be determined below. For the quantity,
$F_i^{\sigma}$, the above boundary condition is written by
\begin{eqnarray}
\label{18}
F_i^{\sigma}=(1-a) \omega_2 \sqrt{3} \frac{c_{\sigma}}{c^2} \frac{\rho_\sigma}{\rho} P_{xy}.
\end{eqnarray}
The particular solution of Eq. (\ref{8a}) is given by the linear combination of the 
decaying exponential solution and the hydrodynamic part. The magnitude of hydrodynamic solution and 
the Knudsen layer part is determined by the boundary condition, Eq. (\ref{18}), in such a way that
\begin{eqnarray}
\label{20}
(1-a) \omega_2 \sqrt{3} \frac{c_{1}}{c^2} \frac{\rho_1}{\rho} P_{xy}&=&
\omega_2 \rho_1 u_s- \omega_2 \rho_1 \frac{\sqrt{3} c_{1}}{\nu_1+\nu_m}\frac{\partial u^{(H)}_y}{\partial x}\nonumber\\
&&+k F^{1(0)}_2,\\
\label{21}
(1-a) \omega_2 \sqrt{3} \frac{c_{2}}{c^2} \frac{\rho_2}{\rho} P_{xy}&=&
\omega_2 \rho_2 u_s- \omega_2 \rho_2 \frac{\sqrt{3} c_{2}}{\nu_1+\nu_m}\frac{\partial u^{(H)}_y}{\partial x}\nonumber\\
&&+k F^{2(0)}_2.
\end{eqnarray}
These equations need to be solved for the unknown quantities, $u_s,k$. It is realized that for obtaining 
the slip velocity, we need the ratio $F^{1(0)}_2/F^{2(0)}_2$ corresponding to the decaying eigenvalue 
$-\lambda_0$. This can be obtained from Eq. (\ref{9}) and the result is given in the Appendix, Eq. (\ref{25}). 

After some straightforward
calculation, we obtain the slip velocity from Eq. (\ref{20}-\ref{21}) as
\begin{eqnarray}
\label{22}
u_s=a B \frac{\mu v_0}{P} \frac{\partial u^{(H)}_y}{\partial x},
\end{eqnarray}
where the obtained value of the $B$ parameter is given by Eq. (\ref{23}) in the Appendix.
It can be seen that we get the desired value of the slip coefficient with the choice $a=\alpha_v/B$.

\begin{figure}
\includegraphics[height=5.5cm,width=8cm]{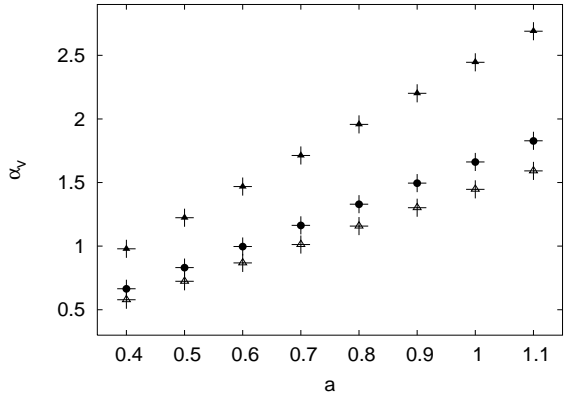}
\caption{\label{}Comparison between the measured slip coefficient obtained from the LB simulation
versus the theoretical derivation at different values of the parameter, $a$. 
$\bullet,\vartriangle,\blacktriangle$ correspond to the values of $m_1/m_2=[1/9.98,1/9.98,1/32.804],
n_1/n=[0.5,0.2,0.6]$, respectively, and represent the results obtained from the LB simulations. $+$ denotes
the analytical results. The collision frequencies are chosen by $v_1=1/(0.05(N-1)),v_m=1/(0.03(N-1))$.
}
\end{figure}

\section{Simulations}

Simulations were performed in order to validate the analytical results obtained above.
The governing equation Eq. (\ref{1}) was solved using the finite difference
method. We mention that the finite difference realization is a useful choice because of the
independence of the velocity vectors and the coordinate grid \cite{ref13,ref13a}. The 
time and the space derivatives are computed using the modified Runge-Kutta and midpoint method, respectively 
\cite{ref13}. {
Note that this scheme is \textit{second order} accurate in both time and space \cite{ref20}.} 
The Kramers problem is simulated in an $N\times1$ domain, $N=128$. The left hand side wall is at rest, 
while the right hand the side moves with a
constant velocity to maintain the constant shear flow in the domain. The proposed
boundary condition with the variable slip coefficient is applied at the left wall,
where the slip coefficient is measured through the relation of Eq. (\ref{15}). Fig. 1 presents
the measured slip coefficient versus the theoretical value obtained from Eq. (\ref{22})
at different values of $a$. Excellent agreement is obtained between the two situations. 
The proposed model can be used to tune the value of the slip coefficient.
Note that above boundary condition can be used in non-stationary case as well.

{
Before concluding our work, we mention that the boundary layer theory can be further
developed involving other flow problems, such as pressure driven flows. In that case, the 
boundary layer equation can be also solved which provides a useful background to find an 
optimal boundary treatment. An another interesting issue is the description of complex 
geometries, slip flow on curved surfaces. For the latter situation, Ref. \cite{ref19}
can be generalized, which describes variable slip coefficient in the single component LB model
on curved boundaries.}

\section{Conclusion}

In this paper, we have developed a new boundary treatment in the LB method for binary
mixtures in order to obtain variable slip coefficient. We have set up a boundary layer
analysis and derived the analytical value of the slip coefficient. Computer simulations
have been performed to validate the model. The results of the simulations are in excellent 
agreement with the analytical derivation. Our method can be used to obtain the desirable
value of the slip coefficient in the lattice Boltzmann model.


\section{Appendix}
The generalized eigenvalue problem, Eq. (\ref{9a}), written out in  components is given by
\begin{eqnarray}
\label{9}
-c_1 \sqrt{3} \lambda F_1^{1(0)}&=&\nu_1 \frac{1}{6} \sum_k F_k^{1(0)} +\nu_m \frac{1}{6} \frac{\rho_1}{\rho} \sum_k (F_k^{1(0)}
+F_k^{2(0)})-(\nu_1+\nu_m)F_1^{1(0)},\nonumber\\
0&=&\nu_1 \frac{2}{3} \sum_k F_k^{1(0)} +\nu_m \frac{2}{3} \frac{\rho_1}{\rho} \sum_k (F_k^{1(0)}+F_k^{2(0)})-(\nu_1+\nu_m)F_0^{1(0)},\nonumber\\
c_1 \sqrt{3} \lambda F_2^{1(0)}&=&\nu_1 \frac{1}{6} \sum_k F_k^{1(0)} +\nu_m \frac{1}{6} \frac{\rho_1}{\rho} \sum_k (F_k^{1(0)}
+F_k^{2(0)})-(\nu_1+\nu_m)F_2^{1(0)},\nonumber\\
-c_2 \sqrt{3} \lambda F_1^{2(0)}&=&\nu_1 \frac{1}{6} \sum_k F_k^{2(0)} +\nu_m \frac{1}{6} \frac{\rho_2}{\rho} \sum_k (F_k^{1(0)}
+F_k^{2(0)})-(\nu_1+\nu_m)F_1^{2(0)},\nonumber\\
0&=&\nu_1 \frac{2}{3} \sum_k F_k^{2(0)} +\nu_m \frac{2}{3} \frac{\rho_2}{\rho} \sum_k (F_k^{1(0)}+F_k^{2(0)})-(\nu_1+\nu_m)F_0^{2(0)},\nonumber\\
c_2 \sqrt{3} \lambda F_2^{2(0)}&=&\nu_1 \frac{1}{6} \sum_k F_k^{2(0)} +\nu_m \frac{1}{6} \frac{\rho_2}{\rho} \sum_k (F_k^{1(0)}
+F_k^{2(0)})-(\nu_1+\nu_m)F_2^{2(0)}.\nonumber\\
\end{eqnarray}

The non-zero eigenvalue of the eigenvalue problem is obtained by
\begin{eqnarray}
\label{10}
\lambda_0=\frac{\sqrt{\nu_m}}{\sqrt{\nu_1+3\nu_m}} \frac{\sqrt{m_1 m_2}}{m} \frac{\nu_1+\nu_m}{c}.
\end{eqnarray}

The ratio of the eigenvalues, $F^{1(0)}_2/F^{2(0)}_2$, corresponding to the decaying eigenvalue $-\lambda_0$
can be obtain from Eq. (\ref{9}) in such a way that
\begin{eqnarray}
\label{25}
F_2^{1(0)}&=&c_2^2(\nu_1+\nu_m+c_1 \sqrt{3} \lambda_0) k,\nonumber\\
F_2^{2(0)}&=&-c_1^2(\nu_1+\nu_m+c_2 \sqrt{3} \lambda_0) k,
\end{eqnarray}
where $k$ is an arbitrary multiplier.

The parameter in the slip velocity, Eq. (\ref{22}), is given in such a way that
\begin{eqnarray}
\label{23}
B=\sqrt{\frac{3}{2}} \frac{n_1 m/ \sqrt{m_1}+n_2 m/ \sqrt{m_2}+n \sqrt{m} A }{n\sqrt{m}+n_1\sqrt{m_1}A+n_2\sqrt{m_2}A},
\end{eqnarray}
with
\begin{eqnarray}
\label{24}
A=\sqrt{3} \frac{\sqrt{v_m}}{\sqrt{v_1+3v_m}}.
\end{eqnarray}

\end{document}